\documentclass[superscriptaddress,twocolumn,aps,pre]{revtex4}
\usepackage{graphicx}
\begin{document}
\title{{\bf Growing Directed Networks: Organization and Dynamics}}
\author{Baosheng Yuan}
\affiliation{Department of Computational Science,
Faculty of Science, National University of Singapore, Singapore 117543}
\author{Kan Chen}
\affiliation{Department of Computational Science, Faculty of Science,
National University of Singapore, Singapore 117543}
\author{Bing-Hong Wang}
\affiliation{Department of Computational Science, Faculty of Science,
National University of Singapore, Singapore 117543}
\affiliation{Department of Modern Physics, University of Science and
 Technology of China, Hefei, Anhui, 230026, China}

\date{\today }

\begin{abstract}

We study the organization and dynamics of growing directed networks.
These networks are built by adding nodes successively in such a way that each new node
has $K$ directed links to the existing ones. The organization of a growing directed network
is analyzed in terms of the number of ``descendants'' of each node in the network.
We show that the distribution $P(S)$ of the size, $S$, of 
the descendant cluster is described generically by a power-law,
 $P(S) \sim S^{-\eta}$, where the exponent $\eta$ depends on the value of $K$ as well 
as the strength of preferential attachment. We determine that, 
in the case of growing random directed networks
 without any preferential attachment, $\eta$ is given by $1+1/K$.
 We also show that the Boolean dynamics of these networks is stable for any value of $K$. However, with a small fraction of reversal in the direction of the links,
the dynamics of growing directed networks appears to operate on ``the edge of chaos'' with a
power-law distribution of the cycle lengths. We suggest that the
growing directed network may serve as another paradigm for 
the emergence of the scale-free features in network organization and dynamics.

\end{abstract}
\maketitle


The dynamics of complex adaptive systems is strongly influenced by
the way the elements of the network are connected and the way
these elements interact. The organization of real-world networks
\cite{Albert, Newman, Dorogovtsev} has attracted intensive interest
following the seminal work of Barab\'asi and Albert \cite{Barabasi} 
on scale-free networks. The degree distribution $P(k)$, which gives the
probability that a randomly selected node has exactly $k$ edges, 
 has been used as the most important characterization of
complex networks. Power-law or scale-free degree distributions
have been found in many real-world complex networks, such as the
Internet, cellular metabolic networks, research collaboration
network, and the World Wide Web \cite{Barabasi, Strogatz}. Most studies of the network have been focused on network topology, but there are also a few studies of dynamical processes on these networks \cite{Anghel,Galstyan,Paczuski}. In particular, Aldana and Cluzel demonstrated that the scale-free topology of the
network favors robust dynamics \cite{Aldana}. Nevertheless, much work is still needed to characterize and classify network dynamics and organization. 

In this paper we study a generic class of growing directed networks, which are grown by adding nodes
successively, just as in the well-known Barab\'asi-Albert model. 
 But we consider the resulting network only as a directed network, and we focus 
on the limiting case of
a citation-like network in which the new node is directed to the
existing nodes and not the other way around. Except for the initial
cluster of $K+1$ nodes, which are linked in both directions, all other
nodes added have $K$ directed links to the existing ones.
In terms of dynamics
the existing nodes are not controlled by the new node added, thus
the network can be viewed essentially as a hierarchical
feed-forward network. To make our discussion relevant to many real
world networks which typically have a fraction of feedback
links, we also consider the modified network that contain a small fraction
of link reversals.

We investigate the global organization of these growing directed
networks. Unlike the undirected version of the network, the
influence of a node on the others in the directed network may be limited.
 For each node we can define a descendant cluster consisting of all the nodes that are
linked to it directly or indirectly through intermediate nodes. More precisely, the descendant cluster of node $v$ is the set of all nodes from which node $v$ can be reached by following a path of directed links. This is the same
as the in-component defined in Ref.~\cite{Krapivsky}. The
possible impact of a given node on the others can be characterized by
the size of its descendant cluster. 
The cluster size distribution gives an
overall description of the network organization in terms of the
potential influence of one node on the others. It is a better measure of the
global organization of the directed network than the degree distribution, which
is essentially a measure of local connectivity in the network. We show that, as far as
the cluster size distribution is concerned, growing directed networks are generically
scale-free, irrespective of the strength of preferential attachment and the value of $K$.

We also investigate the Boolean dynamics of these growing networks.
The study of the  dynamics of Boolean networks was pioneered by Kauffman 
\cite{Kauffman, Kauffman2}, who focused primarily on random directed networks. 
Kauffman's NK model (Kauffman net) consists of N nodes and K directed links per node
such that each node is  controlled by K other nodes. The model
 has been used as a prototypical model
of gene regulation and control. Kauffman suggests that gene
networks operate on the ``edge of chaos'' (in a critical phase), as
evolution demands that there be sensitivity to perturbations and
mutations. In Kauffman's NK model, the critical phase occurs only
at the specific parameter value $K=2$ \cite{Kauffman,Derrida}. For $K>2$
the dynamics is chaotic. We show that, in contrast, the dynamics of growing directed networks (with a small fraction of link reversals) appears to operate on the ``edge of chaos'' for
a wide range of values of $K$.

Two models of growing directed networks are considered in our study. In
Model A, the network is grown in the same way as in the undirected
version \cite{Krapivsky2}. We start with an initial cluster of $K+1$ nodes,
which are fully connected (two directed links between
each pairs of nodes). At each stage, we add a new node with $K$ links
to $K$ of the nodes already present in the network. The link is
directed from the new node to an existing node, meaning that the
new node is a descendant of the existing one.  We assume that the probability 
of connecting a new node to an existing one with degree $k$ is proportional to
$k^{\alpha}$, where $k$ is the total degree of the node: $k=k_{in}+k_{out}$ 
($k_{in}$ is the number of incoming links and $k_{out}$ is the number of outgoing links; for
this model $k_{out} \equiv K$). For $\alpha > 0$, we have
preferential attachment. The undirected version of the network
with $\alpha=1$ corresponds to the Barab\'asi-Albert model. The
power-law degree distribution can only be observed for $\alpha =
1$ \cite{Krapivsky2, Krapivsky3}. As far as the degree distribution is concerned,
the network is only scale-free when $\alpha = 1$. In contrast, we found that
the power-law distribution is in fact quite generic in the size
distribution of descendant clusters. Fig.~1 shows such size distribution for
$\alpha=0,0.5,1.0$ and $K=5$. 
The degree distributions are also shown in the figure for comparison.
For $\alpha \le 1.0$ the cluster size distribution  can be
described very well by a power law $P(S)\sim S^{-\eta}$ (The deviation is
noticeable only when $S \sim O(N)$). For $\alpha=0$ and $0.5$, we found
$\eta \approx 1.2$, and for $\alpha =1$, $\eta \approx 1.3$. 
We have checked that the power laws exist for a
wide range of values of $K$; the exponent depends on
both $K$ and $\alpha$. Even though $\alpha =1$ is a special case
for the degree distribution (as can also be seen from Fig.~1), it is
not for the descendant cluster distribution. The power-law distribution of
descendant cluster sizes is rather generic in growing directed network.

\begin{figure}
\includegraphics*[width=8.5cm]{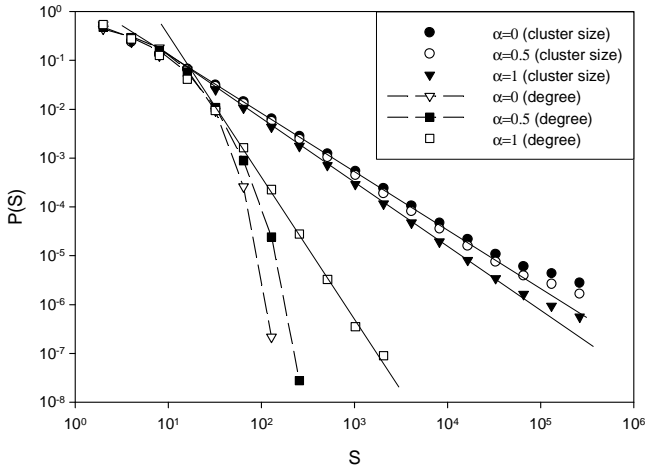}
\caption{ Model A: The descendant cluster size distribution for $\alpha=0, 0.5, 1$.
 $N=200000$ and $K=5$ are used.}
\end{figure}

We also consider a variant of Model A, which we refer to as Model B. 
In this model, we
choose the growth rule such that the probability of connecting to
the node with in-degree $k_{in}$ is proportional to
$k_{in}^{\alpha} + 1$. The constant $1$ is added to give a nonzero
starting weight to the nodes that have not been connected to. Again we obtain power-law cluster size
distributions, which are plotted in Fig.~2. The exponents obtained
for the cluster distribution are different from those of Model A.
We obtained $\eta \approx 1.2$ for $\alpha=0$, $\eta \approx 1.3$
for $\alpha =0.5$, and $\eta \approx 1.6$ for $\alpha=1$.

\begin{figure}
\includegraphics*[width=8.5cm]{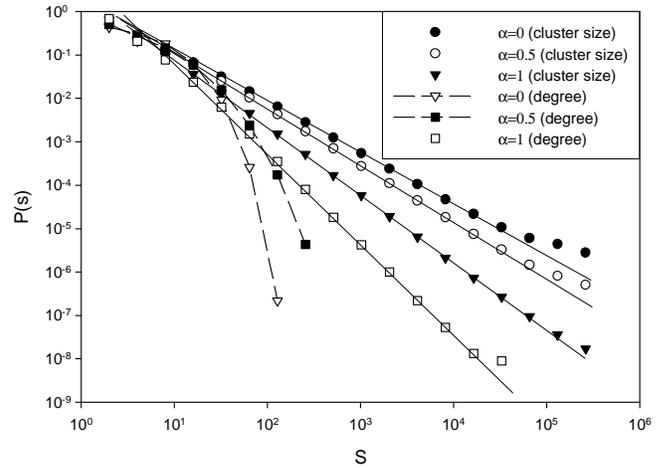}
\caption{ Model B: The descendant cluster size distribution for 
$\alpha=0, 0.5, 1$. $N=200000$ and $K=5$ are used.}
\end{figure}

Model A and B are the same model when $\alpha =0$ (no preferential
attachment). For this case we can write down a Master equation for
the cluster size distribution. Let $n(N, S)$ be the number of
clusters of size $S$ when $N$ nodes are present. Now we add a new
node to the network. For $N\gg K$, $n(N,S)$ evolves according to
the following equation:
\begin{eqnarray}
n(N+1, S+1)=&&n(N,S)\left[1-(1-\frac{S}{N})^K\right] \nonumber \\
        &&+ n(N,S+1)(1-\frac{S+1}{N})^K,
\end{eqnarray}
where $(1-\frac{S}{N})^K$ is the probability that the new node
added does not link to any of the $S$ nodes in a given cluster of
size $S$. For $1\ll S\ll N$, $n(N,S)$ can be approximated as
$n(N,S) = N*p(s)$, where $s=S/N$ and $p(s)$ is the probability density
function for the size distribution. In addition,
$1-(1-\frac{S}{N})^K \approx KS/N=Ks$. In terms of $p(s)$ the
above equation can be rewritten as
\begin{equation}
\label{ps}
p(s+\frac{1}{N}) = -N\left[K(s+\frac{1}{N})p(s+\frac{1}{N})-Ksp(s)\right]
\end{equation}
Neglecting the terms of order $1/N$ and higher, we have
\begin{equation}
Ks\frac{dp}{ds} = -(K+1)p(s)
\end{equation}
This leads to  $p(s) \propto s^{-\eta}$ with $\eta = 1+1/K$.
Fig.~3 shows the cluster size distribution for $\alpha = 0$ and
$K=1,2,3,5,8$. The numerical values of the exponent $\eta$ agree
very well with the analytical values.

\begin{figure}
\includegraphics*[width=8.5cm]{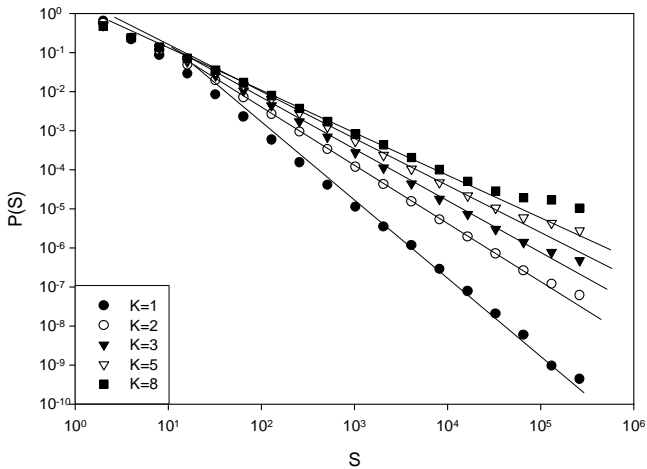}
\caption{The cluster size distribution for $\alpha=0$,
$K=1,2,3,5,8$, and $N=200000$. The slopes of the lines drawn are
given by $-\frac{1+K}{K}$}
\end{figure}

The cluster distribution for the special case of $\alpha=0$ and $K=1$ was first
obtained by Krapivsky and Redner \cite{Krapivsky}. They obtained $\eta = 2$
in agreement with our analysis. They also pointed out that 
the distribution for the special case of $\alpha=1$ and $K=1$ can
also be obtained analytically, and $\eta$ is again equal to 2. This is not
surprising, for both cases, the probability that a new node is added to a cluster
of size S is simply $S/N$. This can be seen from the fact that, for $\alpha =1$,
 this probability is proportional to  $\sum k_{in}/N$, which is the same as $S/N$, because
$\sum k_{in} = \sum k_{out} =S$ for the cluster.
 Thus exactly same Master equation applies for both cases.
We have also checked, numerically, that $\eta$ is in fact equal to 2 for $0\le \alpha \le 1$.

We can also generalize the model to the case $K<1$. The meaning of $K$ for this case is the probability that the new node gets connected to a randomly chosen existing node (the resulting network consists of disconnected components). Let us consider again the special case of $\alpha=0$. It is easy to check that the same Master equation (Eq.~\ref{ps}) can be used to determine $p(s)$, leading again to $p(s) \propto s^{-\eta}$ with $\eta=1+1/K > 2$. It is interesting to note the component size distribution of growing (undirected) networks in the subcritical regime follows exactly the same scaling (see Eq.~(31) of Ref.~\cite{Krapivsky4}, which contains a review of universal properties of growing networks and earlier references on this subject). This is not surprising as the same Master equation applies to both the distribution of the descendant cluster and that of the component size. However, there is no direct one-to-one mapping which maps a descendant cluster to a component of the network (each node has a descendant cluster, but the number of connected components is typically much less than the number of nodes).

We now turn to the Boolean dynamics of growing directed networks.
The study of the dynamics of Boolean networks was pioneered by
Kauffman, who used it as a prototypical model of a genetic
regulatory network. In the Kauffman net, each node is controlled by
K other nodes chosen randomly. The dynamics of the
Kauffman net depends crucially on $K$. For $K=1$ the dynamics
converges to a fixed point or a limit cycle (this is the ordered
phase). For $K=2$, the system is at the ``edge of chaos" where
cycles of many different lengths can appear. For $K>2$, the system
is in the chaotic phase.

\begin{figure}
\includegraphics*[width=8.0cm]{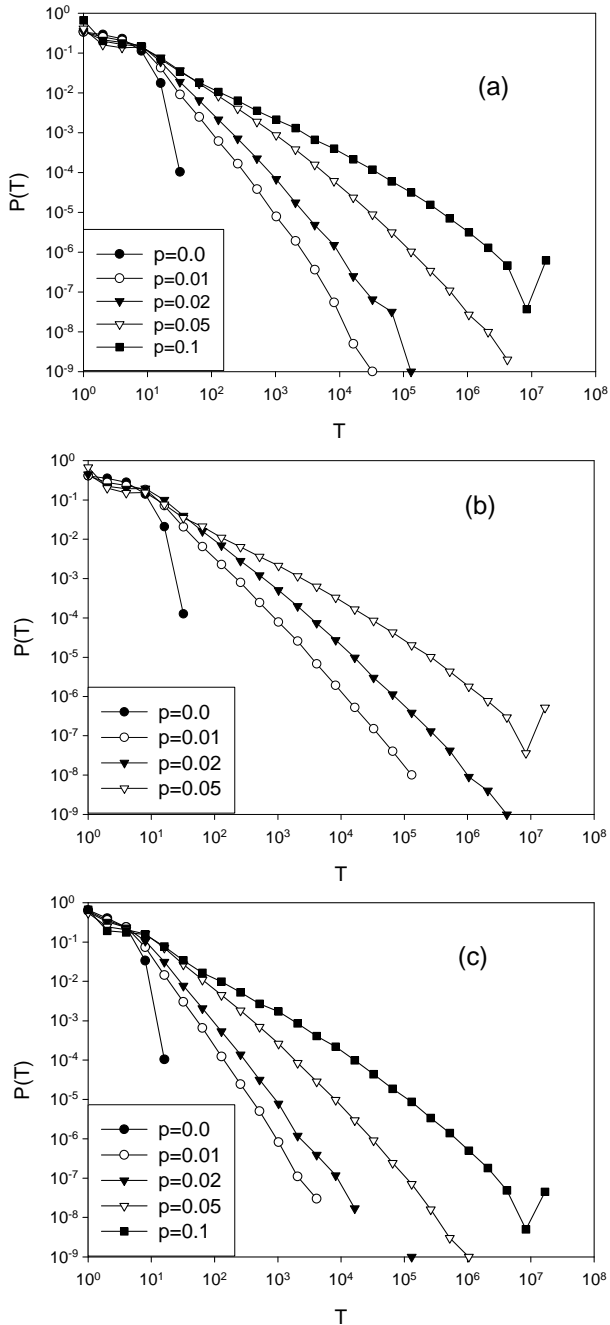}
\caption{The distribution of cycle lengths in growing random directed networks ($\alpha=0$) for the cases: (a) $K=6$, $N=401$, (b) $K=6$, $N=801$, and (c) $K=4$, $N=801$. The distribution (for $0<p<p_c$)  can be approximated by a power law $P(T) \sim T^{-\theta}$}
\end{figure}

The Boolean dynamics of our growing directed network is always in
the ordered phase with the maximum period equal to $2^{K+1}$. This is
due to the fact that, by construction, the initial $K+1$ nodes are
mutually connected; they are not controlled by the nodes added to
the network later. Thus the maximum period for the dynamics of this
initial cluster is $2^{K+1}$. As the new nodes are controlled only
by the existing nodes in the cluster, it is easy to show, by induction,
that the period of the dynamics of the entire system is the same
as the period of the dynamics of the initial cluster. Thus the
dynamics of the growing directed network is in the ordered
phase irrespective of the values of $K$ and $\alpha$.

\begin{figure}
\includegraphics*[width=8.0cm]{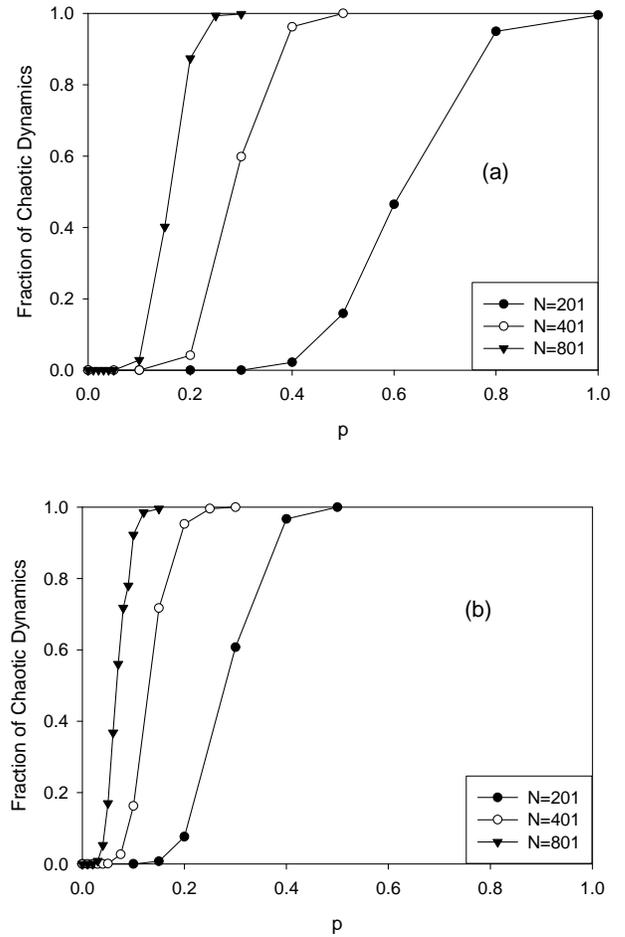}
\caption{The fraction of chaotic dynamics found in growing random directed networks as a function of $p$ for $N=201, 401, 801$,  and (a) $K=4$ and (b) $K=6$.} 
\end{figure}

Many real-world networks, with the possible exception of
citation-like networks, are not simple feed-forward directed
networks we considered above. There typically exist a certain
fraction of feedback links (the new node controls the existing one). 
To model these networks, we
start with the original feed-forward network. Then, with a
probability $q=p/K$ (Here $p$ is the probability that a node has a feedback link), 
reverse the direction of the links. Here we focus on the case of $\alpha=0$ (growing random directed networks). For sufficiently small value of $p$,
the Boolean dynamics of these modified networks is similar to that of the
Kauffman net when K=2 \cite{Bhattacharjya}: Depending on the initial conditions, cycles
of a wide range of lengths appear. We performed extensive simulation of the Boolean dynamics of these networks for various values of $\alpha$, $K$, and $N$. The length of the simulation is up to $10^7$ time steps so that we can detect the period of the dynamics up to the order of $10^7$. The statistical results were obtained by averaging over 2000 to 200000 independent runs.  Fig.~4 shows the distribution of the cycle
length (period), which can be described as a power law $P(T) \sim T^{-\theta}$,  where the exponent increases as $p$ increases. We also found that there is a threshold value $p_c$, which depends on both $K$ and $N$. The power law distribution of $P(T)$ occurs for $p<p_c$, with the exponent $\theta$ approaches $1$ as $p\rightarrow p_c$. For $p > p_c$, the fraction of chaotic dynamics (numerically we classify the dynamics as chaotic if the period is greater than $10^7$) increases rapidly. There is a clear transition from the ``edge of chaos" regime to a chaotic regime. Fig.~5 shows the fraction of chaotic dynamics as a function of $p$, for $K=4$ and $6$, $N=201, 401$, and $801$. The transition is rather sharp for large $N$. We have also checked that the same qualitative behaviors can be observed for small $\alpha$ ($\alpha<0.5$). However, for $\alpha$ close to $1$, even though there is still a transition to a chaotic regime, there is no ``edge of chaos"' regime for $p<p_c$ (the distribution $P(T)$ is  broad, but it cannot be described using a power law).  This shows that the dynamical properties of the network are not necessarily correlated with the local degree distribution. It remains a challenging task, however, to identify the key features of the global organization of the networks that directly affect the dynamics.

In conclusion, we have studied the organization and Boolean
dynamics of growing directed networks. In terms of cluster of
descendants, the size distribution exhibits a robust power law for
a wide range of $K$ and $\alpha$ values.  The Boolean dynamics of
the networks  is very stable with the maximum period equal to
$2^{K+1}$. However, with a small fraction of link reversals, the
dynamics appears to operate on the ``edge of chaos'' with a
power-law distribution of the cycle lengths. This critical regime
is rather generic and can be obtained without a fine tuning of the
parameters $K$ and $\alpha$, in contrast to the original Kauffman model. With its generic scale-free features in the
organization and dynamics, the growing directed network  serves as
another paradigm for the emergence of scale-free dynamical and organizational
properties as exhibited in many real-world networks.

We thank C. Jayaprakash for careful reading of the manuscript and very helpful comments. 
This work is supported by the National University of Singapore
research grant R-151-000-028-112. BHW also acknowledges the
support by the National Natural Science Foundation of China
(No.70271070), and by the Specialized Research Fund for the
Doctoral Program of Higher Education (No. 20020358009).

\end{document}